\documentclass[aps,pre,reprint,showpacs,superscriptaddress]{revtex4-1}
\pdfoutput=1

\usepackage{graphicx}
\usepackage{subfigure}
\usepackage{color}
\usepackage{amsmath}
\usepackage{amssymb}
\usepackage{epstopdf}
\usepackage[linktocpage,colorlinks=true,linkcolor=blue,citecolor=blue,breaklinks=true]{hyperref}
\usepackage{verbatim}
\usepackage{breakcites}
\usepackage[version=3]{mhchem}

\newcommand{\be}{\begin{equation}}
\newcommand{\ee}{\end{equation}}

\begin{document}

\preprint{}

\title{Maximal stochastic transport in the Lorenz equations}

\author{Sahil Agarwal}
\affiliation{Program in Applied Mathematics, Yale University, New Haven, USA}
\affiliation{Mathematical Institute, University of Oxford, Oxford, UK}

\author{J. S. Wettlaufer}
\affiliation{Program in Applied Mathematics, Yale University, New Haven, USA}
\affiliation{Mathematical Institute, University of Oxford, Oxford, UK}
\affiliation{Nordita, Royal Institute of Technology and Stockholm University, SE-10691 Stockholm, Sweden}
\email[]{john.wettlaufer@yale.edu}

\date{\today}

\begin{abstract}
We calculate the stochastic upper bounds for the Lorenz equations using an extension of the background method.  
In analogy with Rayleigh-B\'enard convection the upper bounds are for heat transport versus Rayleigh number.  As
might be expected, the stochastic upper bounds are larger than the deterministic counterpart of \citet{Doering15}, 
but their variation with noise amplitude exhibits interesting behavior.  Below the transition to chaotic 
dynamics the upper bounds increase monotonically with noise amplitude.  However, in the chaotic regime this 
monotonicity depends on the number of realizations in the ensemble; at a particular Rayleigh number the bound may increase or decrease with noise amplitude.  
The origin of this behavior is the coupling between the noise and unstable periodic orbits, the degree of which depends on the degree to which the ensemble represents the ergodic set.   This is confirmed by examining the close returns plots of the full solutions to the stochastic equations and the numerical convergence of the noise correlations.  The numerical convergence of both the ensemble and time averages of the noise correlations is sufficiently slow that it is the limiting aspect of the realization of these bounds.  Finally, we note that the full solutions of the stochastic equations demonstrate that the effect of noise is equivalent to the effect of chaos.
\end{abstract}


\maketitle

\section{Introduction} 
Noise is an integral part of any physical system. It can be ascribed to fluctuations arising from intermittent forcing, observational uncertainties, interference from external sources or unresolved physics. In circumstances where noise acts to destroy a signal of interest, it is viewed as a nuisance.  However, it can also be the case that fluctuations act to stabilize a system, examples of which include noise-induced optical multi-stability \cite{Barberoshie93}, asymmetric double well potentials \cite{Mankin08}, plant ecosystems \cite{Odorico05}, population dynamics \cite{Parker2011}, and in electron-electron interactions in quantum systems \cite{Landauer:1998aa}.
Curiously, it has recently been shown that noise can have positive effects on cognitive functions such as learning and memory \cite{Rausch:2013aa}.  Finally, a key issue arising when examining observational data is whether fluctuations are intrinsic or due to external forcing, which can be confounded by temporal multifractality \cite[e.g.,][]{Sahil:MF}.

Given the breadth of settings in which the effects of noise manifest themselves on dynamical systems, it appears prudent to examine such matters in a well studied and yet broadly relevant system.  Thus, we study the influence of noise in the Lorenz system \cite{Lorenz63}, which is an archetype of deterministic nonlinear dynamics.  Moreover, \citet{Doering15} have recently determined the maximal (upper bounds) transport in the Lorenz equations, thereby providing us with a rigorous test bed for stochastic extensions.   In \S \ref{sec:Lorenz} we describe the stochastic Lorenz model, followed by the derivation of the stochastic upper bounds in \S \ref{sec:SUB}.   We interpret the core results and their implications in \S \ref{sec:Results} before concluding.  

\section{Stochastic Lorenz Model\label{sec:Lorenz}} 

The Lorenz model is a Galerkin-modal truncation of the equations for Rayleigh-B\'enard convection with stress-free boundary conditions on the upper and lower boundaries.  
It acts as a rich toy model of low-dimensional chaos and since it's origin extensive studies have been made spanning a wide range of areas \cite[e.g.,][]{Strogatz:2014}.  
Of particular relevance here, is using the system as a model for heat transport in high Rayleigh number turbulent convection \cite{Doering15}.  

%

The stochastic form of the Lorenz system is described by the following coupled nonlinear ordinary differential equations,
\begin{eqnarray}
\label{eq:StoLorenz}
\frac{d}{dt}X &=& \sigma (Y - X) + A_{1}\xi_{1}, \nonumber \\
\frac{d}{dt}Y &=& X(\rho - Z) - Y + A_{2}\xi_{2},  \\
\frac{d}{dt}Z &=& XY - \beta Z + A_{3}\xi_{3} \nonumber
\end{eqnarray}
where $X$ describes the intensity of convective motion, $Y$ the temperature difference between ascending and descending flow and $Z$ the deviation from linearity of the vertical temperature profile. The control parameters are $\sigma$ the \emph{Prandtl Number}, $\rho$ the \emph{Rayleigh Number} and $\beta$ a domain geometric factor. The $A_{i}$ are the noise amplitudes and $\xi_{i}$ are the noise processes. Clearly, the deterministic system has $A_{i} = 0$. 

This type of additive noise may appear, for example, in observational errors, when the errors do not depend on the system state or as a model of sub-grid scale processes approximated by noise associated with unexplained physics \cite{Arnold:2013aa}. In multiplicative noise the system has an explicitly state dependent noise process. 

Although real noise will always have a finite time correlation, taking the limit that the noise correlation goes to zero as $\Delta t \to 0$, serves as a good approximation for the noise forcing. This is the \emph{white noise} limit of colored noise forcing.  White noise forcing $\xi(t)$ is defined by an autocorrelation function written as
\begin{equation}
  \langle \xi(t)\xi(s) \rangle = 2D\delta(t-s),
\label{eq:white}
\end{equation}
where, $t-s$ is the time lag, $D$ is the amplitude of the noise, $\langle\bullet\rangle$ represents the time average and $\delta(r)$ is the Dirac delta-function.

\section{Stochastic Maximal Transport\label{sec:SUB}} 
Initiated by the work of Louis Howard \cite{Howard:1972}, maximizing the transport of a quantity such as heat or mass is a core organizing principle in modern studies of dissipative systems. In this spirit
\citet{Doering15} studied the transport in the deterministic Lorenz equations and determined 
the upper bound, which depends on the exact steady solutions $X_s$, $Y_s$, as $\overline{\lim_{T \rightarrow \infty}} \left \langle XY \right \rangle _{T}  = X_s Y_s = \beta (\rho - 1)$, where $X_s = Y_s = \pm \sqrt{\beta (\rho - 1)}$ for $\rho \geq 1$. Moreover, they showed that any time-dependent forcing would decrease the transport in the system, and hence the steady state maximizes the transport in the system. We study the effect of noise on the maximal transport in this system as the Rayleigh number $\rho$ is varied. 

Let $X = x, Y = \rho y, Z = \rho z$ and $A_{1} = A_{2} = A_{3} = A$ in the system of equations \ref{eq:StoLorenz}, which transform to 
\begin{eqnarray}
\label{eq:StoLorenzTransform}
\frac{d}{dt}x &=& \sigma (\rho y - x) + A \xi_{1}, \nonumber \\
\frac{d}{dt}y &=& x(1 - z) - y + \frac{A}{\rho}\xi_{2},  \\
\frac{d}{dt}z &=& xy - \beta z + \frac{A}{\rho}\xi_{3}. \nonumber
\end{eqnarray}

In the next two sub-sections, we calculate the stochastic upper bound of equations \ref{eq:StoLorenzTransform} using both \emph{It\^{o}} and \emph{Stratonovich} calculi.

\subsection{It\^{o} Calculus Framework}

Now, knowing that the state variables ($x,y,z$) in the Lorenz system are bounded \cite{Doering15, Doering95}, and following the approach of \citet{Doering15} for this stochastic system, the long time averages of $\frac{1}{2}x^2, \frac{1}{2}(y^2 + z^2)$ and $-z$ can be written as
\begin{equation}
0 = -\langle x^2 \rangle _{T} + \rho \langle xy \rangle _{T} + \frac{A^2}{2\sigma} + \frac{A}{\sigma}\langle x \xi_{1} \rangle _{T} + O(T_{-1}),
\label{eq:IXav}
\end{equation}
\begin{align}
0 = &-\langle y^2 \rangle _{T} + \langle xy \rangle _{T} - \beta \langle z^2 \rangle _{T} + \frac{A^2}{\rho ^2} + \frac{A}{\rho}\langle y \xi_{2} \rangle _{T} \nonumber \\
&+ \frac{A}{\rho}\langle z \xi_{3} \rangle _{T}
+ O(T_{-1}),
\label{eq:IYav}
\end{align}
\begin{equation}
0 = -\langle xy \rangle _{T} + \beta \langle z \rangle _{T} + O(T_{-1}),
\label{eq:IZav}
\end{equation}
where, the terms $\frac{A^2}{2\sigma}$ in Eq. \ref{eq:IXav} and $\frac{A^2}{\rho ^2}$ in Eq. \ref{eq:IYav} are a consequence of \emph{It\^{o}'s lemma}.

Now, let $z = z_0 + \lambda (t)$, where $z_0 = \frac{r-1}{r}$ is time-independent \cite{Doering15}, and equations \ref{eq:IYav} and \ref{eq:IZav} now become,
\begin{align}
0  =& -\langle y^2 \rangle _{T} + \langle xy \rangle _{T} - \beta z_{0}^{2} - 2\beta z_0 \langle \lambda \rangle _{t} - \beta \langle \lambda^{2} \rangle _{T} \nonumber \\ 
     & + \frac{A^2}{\rho ^2} + \frac{A}{\rho}\langle y \xi_{2} \rangle _{T} + \frac{A}{\rho}\langle \lambda \xi_{3} \rangle _{T} + O(T_{-1}),~ \text{and}
\label{eq:IYav1}
\end{align}
\begin{align}
0 = -\langle xy \rangle _{T} + \beta  z_0  + \beta \langle \lambda \rangle _{T} + O(T_{-1}).
\label{eq:IZav1}
\end{align}
Therefore, equation (\ref{eq:IYav1}) $ + 2 z_0 \times $ (\ref{eq:IZav1}) becomes,
\begin{align}
0 =& -\langle y^2 \rangle _{T} + (1 - 2 z_0) \langle xy \rangle _{T} + \beta z_{0}^{2} - \beta \langle \lambda^{2} \rangle _{T} \nonumber \\
   & + \frac{A^2}{\rho ^2} + \frac{A}{\rho}\langle y \xi_{2} \rangle _{T} + \frac{A}{\rho}\langle \lambda \xi_{3} \rangle _{T} + O(T_{-1}). 
\label{eq:IYZav}
\end{align}
Now adding $\frac{1}{\rho} \times$ (\ref{eq:IXav}) to $\rho \times$ (\ref{eq:IYZav}) gives
\begin{align}
0 =& -\rho \langle y^2 \rangle _{T} + \rho (1 - 2 z_0) \langle xy \rangle _{T} + \rho \beta z_{0}^{2} - \rho \beta \langle \lambda^{2} \rangle _{T}  \nonumber \\
& - \frac{1}{\rho} \langle x^2 \rangle _{T} + \langle xy \rangle _{T} + \frac{A}{\rho \sigma}\langle x \xi_{1} \rangle _{T} + \frac{A^2}{\rho} \nonumber \\
     &  + \frac{A^2}{2\rho \sigma} + A \langle y \xi_{2} \rangle _{T} + A \langle \lambda \xi_{3} \rangle _{T} + O(T_{-1}),
\label{eq:IXYZav}
\end{align}
and adding $(\rho - 1)\langle xy \rangle _{T}$ to both sides gives, 
\begin{align}
(\rho -1)\left \langle xy \right \rangle _{T} = & \rho \beta z_o^2 + A\left [ \langle y \xi_2 \rangle _{T} + \langle \lambda \xi_3 \rangle _{T} + \frac{1}{\sigma \rho}\langle x \xi_1 \rangle_{T} \right ] \nonumber \\
& - \left \langle \left ( \frac{x}{\sqrt{\rho}} - \sqrt{\rho}y \right )^2 + \rho \beta \lambda^2 \right \rangle _{T} \nonumber \\
& + A^2 \left[ \frac{1}{\rho} + \frac{1}{2\rho \sigma} \right ] + O(T^{-1}).
\label{eq:IXYZ}
\end{align}
We thus arrive at 
 \begin{eqnarray}
 (\rho -1)\left \langle xy \right \rangle _{T} &\leq&  \rho \beta z_o^2 + A^2 \left[ \frac{1}{\rho} + \frac{1}{2\rho \sigma} \right ] \nonumber \\
 & &+ A\left[\langle y \xi_2 \rangle _{T} + \langle \lambda \xi_3 \rangle _{T} + \frac{1}{\sigma \rho}\langle x \xi_1 \rangle_{T}\right] \nonumber \\
 & & + O(T^{-1}).
\label{eq:IXYZleq}
\end{eqnarray}
Comparing equation \ref{eq:IXYZleq} above with equation $19$ from \citet{Doering15}, we see an additional term due to the stochastic forcing
\begin{eqnarray}
\overline{\lim_{T \rightarrow \infty}} \left \langle XY \right \rangle _{T} &=& \overline{\lim_{T \rightarrow \infty}} \rho\left \langle xy \right \rangle _{T} \nonumber \\
&\leq& \beta(\rho - 1) + \frac{A^2}{\rho - 1} \left[ 1 + \frac{1}{2\sigma} \right ] \nonumber \\
& & \hspace{-15 mm} + \frac{\rho A}{\rho - 1} \left [\frac{1}{\rho}\langle Y \xi_2 \rangle _{T} + \langle \lambda \xi_3 \rangle _{T} + \frac{1}{\sigma \rho}\langle X \xi_1 \rangle_{T} \right ], 
\label{eq:IXYZlimit}
\end{eqnarray}
which shows that the stochastic upper bound transcends the deterministic upper bound.

\subsection{Stratonovich Calculus Framework}

\begin{figure}[h]
\includegraphics[trim = 20 0 30 20, clip, width = 1\linewidth]{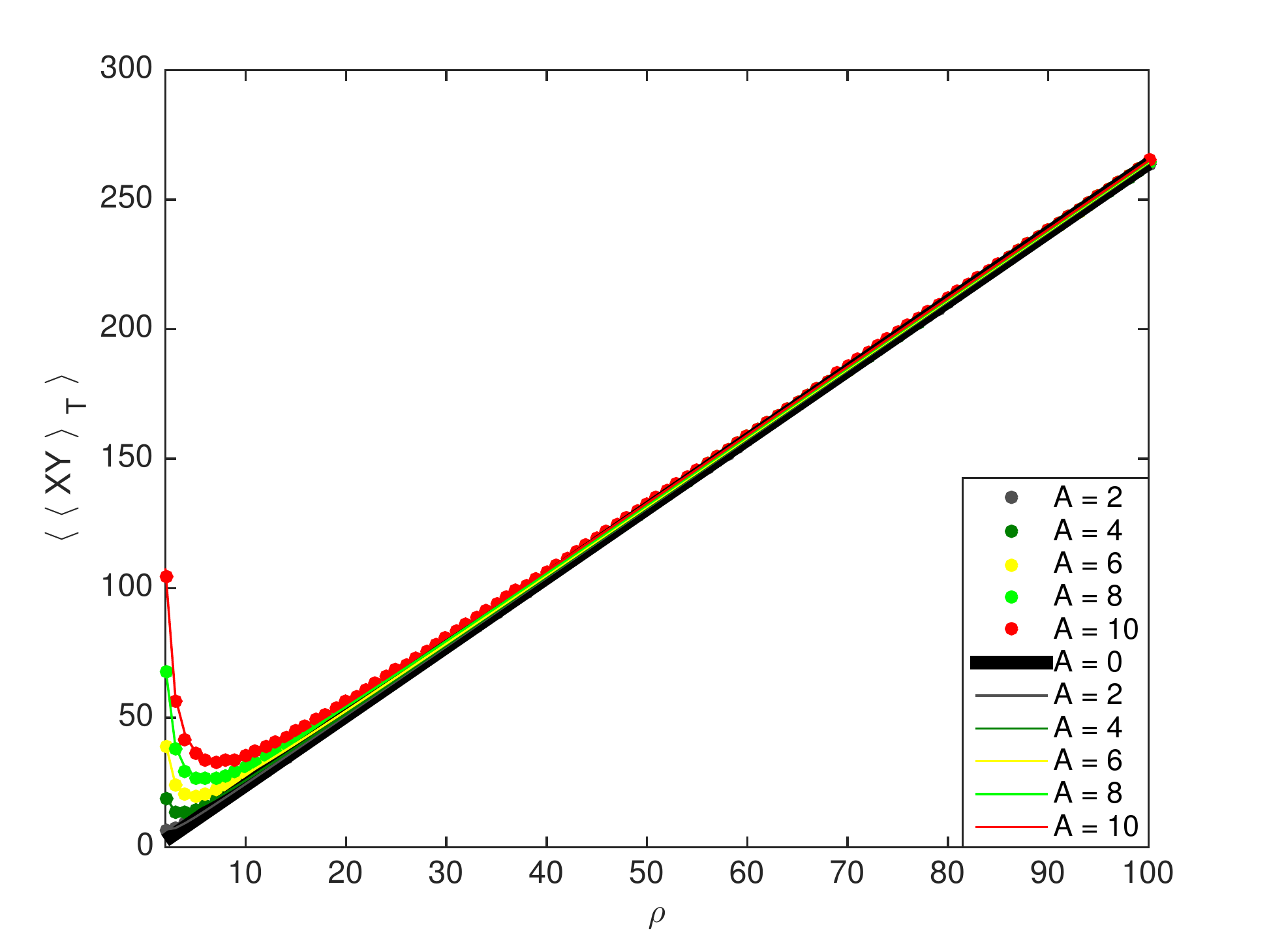}
\vspace{-0.3in}
\caption{$\langle \overline{\lim_{T \rightarrow \infty}} \left \langle XY \right \rangle _{T} \rangle$ as a function of $\rho$ and noise amplitude $A$, with black line showing the upper bound in the deterministic case \cite{Doering15}, colored lines showing the analytical solution from equation \ref{eq:XYZleq} and solid circles showing the numerical solution as the ensemble average in equation \ref{eq:SXYZlimit}.} 
\label{fig:SUBXY}
\vspace{-0.2in}
\end{figure}

In this framework, the equations analogous to \ref{eq:IXav}, \ref{eq:IYav} and \ref{eq:IZav} are 
\begin{equation}
0 = -\langle x^2 \rangle _{T} + \rho \langle xy \rangle _{T} + \frac{A}{\sigma}\langle x \xi_{1} \rangle _{T} + O(T_{-1}),
\label{eq:SXav}
\end{equation}
\begin{align}
0 =& -\langle y^2 \rangle _{T} + \langle xy \rangle _{T} - \beta \langle z^2 \rangle _{T} + \frac{A}{\rho}\langle y \xi_{2} \rangle _{T} \nonumber \\
&+ \frac{A}{\rho}\langle z \xi_{3} \rangle _{T} + O(T_{-1}),
\label{eq:SYav}
\end{align}
\begin{equation}
0 = -\langle xy \rangle _{T} + \beta \langle z \rangle _{T} + O(T_{-1}),
\label{eq:SZav}
\end{equation}
Again letting $z = z_0 + \lambda (t)$, where $z_0 = \frac{r-1}{r}$, equations \ref{eq:SYav} and \ref{eq:SZav} now become,
\begin{eqnarray}
0  =& -\langle y^2 \rangle _{T} + \langle xy \rangle _{T} - \beta z_{0}^{2} - 2\beta z_0 \langle \lambda \rangle _{t} - \beta \langle \lambda^{2} \rangle _{T} \nonumber \\ 
     & + \frac{A}{\rho}\langle y \xi_{2} \rangle _{T} + \frac{A}{\rho}\langle \lambda \xi_{3} \rangle _{T} + O(T_{-1}),
\label{eq:SYav1}
\end{eqnarray}

\begin{equation}
0 = -\langle xy \rangle _{T} + \beta  z_0  + \beta \langle \lambda \rangle _{T} + O(T_{-1}),
\label{eq:SZav1}
\end{equation}
and hence \ref{eq:SYav1} $+ 2 z_0 \times $ \ref{eq:SZav1} becomes,
\begin{eqnarray}
0 =& -\langle y^2 \rangle _{T} + (1 - 2 z_0) \langle xy \rangle _{T} + \beta z_{0}^{2} - \beta \langle \lambda^{2} \rangle _{T} \nonumber \\
   & + \frac{A}{\rho}\langle y \xi_{2} \rangle _{T} + \frac{A}{\rho}\langle \lambda \xi_{3} \rangle _{T} + O(T_{-1}). 
\label{eq:SYZav}
\end{eqnarray}
Now adding $\frac{1}{\rho} \times $ (\ref{eq:SXav}) to $\rho \times $ (\ref{eq:SYZav}) we find
\begin{eqnarray}
0 =& -\rho \langle y^2 \rangle _{T} + \rho (1 - 2 z_0) \langle xy \rangle _{T} + \rho \beta z_{0}^{2} - \rho \beta \langle \lambda^{2} \rangle _{T}  - \frac{1}{\rho} \langle x^2 \rangle _{T} \nonumber \\
     & + \langle xy \rangle _{T} + \frac{A}{\rho \sigma}\langle x \xi_{1} \rangle _{T} \nonumber \\
     & + A \langle y \xi_{2} \rangle _{T} + A \langle \lambda \xi_{3} \rangle _{T} + O(T_{-1}). 
\label{eq:SXYZav}
\end{eqnarray}
Finally, adding $(\rho - 1)\langle xy \rangle _{T}$ to both sides gives
\begin{align}
(\rho -1)\left \langle xy \right \rangle _{T} = & \rho \beta z_o^2 + A\left [ \langle y \xi_2 \rangle _{T} + \langle \lambda \xi_3 \rangle _{T} + \frac{1}{\sigma \rho}\langle x \xi_1 \rangle_{T} \right ] \nonumber \\
& - \left \langle \left ( \frac{x}{\sqrt{\rho}} - \sqrt{\rho}y \right )^2 + \rho \beta \lambda^2 \right \rangle _{T} \nonumber \\
& + O(T^{-1}).
\label{eq:SXYZ}
\end{align}
We thus arrive at
 \begin{eqnarray}
 (\rho -1)\left \langle xy \right \rangle _{T} &\leq&  \rho \beta z_o^2 \nonumber \\
 & & + A\left [\langle y \xi_2 \rangle _{T} + \langle \lambda \xi_3 \rangle _{T} + \frac{1}{\sigma \rho}\langle x \xi_1 \rangle_{T} \right ] \nonumber \\
 & & + O(T^{-1})
\label{eq:SXYZleq}
\end{eqnarray}
Now, comparing Eq. \ref{eq:SXYZleq} above with Eq. $19$ from \citet{Doering15}, we see an additional term due to the stochastic forcing, which, as expected from the previous section, increases the upper bound; 
\begin{eqnarray}
\overline{\lim_{T \rightarrow \infty}} \left \langle XY \right \rangle _{T} &=& \overline{\lim_{T \rightarrow \infty}} \rho\left \langle xy \right \rangle _{T} \nonumber \\
&\leq& \beta(\rho - 1) \nonumber \\
& &\hspace{-15 mm} + \frac{\rho A}{\rho - 1} \left [ \frac{1}{\rho}\langle Y \xi_2 \rangle _{T} + \langle \lambda \xi_3 \rangle _{T} + \frac{1}{\sigma \rho}\langle X \xi_1 \rangle_{T} \right ]
\label{eq:SXYZlimit}
\end{eqnarray}
Due to the fact that the noise is additive, the upper-bounds from \emph{It\^{o}} and \emph{Stratonovich} calculi should be equivalent.  This is indeed the case because for the
\emph{It\^{o}} result, $\langle X\xi_1 \rangle = \langle Y\xi_2 \rangle = \langle \lambda \xi_3 \rangle = 0$, whereas for the \emph{Stratonovich} result $\langle X\xi_1 \rangle = \langle Y\xi_2 \rangle = \frac{A}{2}$ and $\langle \lambda\xi_3 \rangle = \frac{A}{2\rho}$. Therefore, when we take the ensemble average of equations \ref{eq:SXYZlimit} and \ref{eq:IXYZlimit} we obtain
\begin{equation}
\langle \overline{\lim_{T \rightarrow \infty}} \left \langle XY \right \rangle _{T} \rangle \leq  \beta(\rho - 1) + \frac{A^2}{\rho - 1} \left[ 1 + \frac{1}{2\sigma} \right ]\label{eq:XYZleq}
\end{equation}
We plot equation \ref{eq:XYZleq} in Fig. \ref{fig:SUBXY}, wherein the lines show the analytic solution and the solid circles denote the numerical solution, taking the ensemble average of equation \ref{eq:SXYZlimit}, all as a function of noise amplitude $A$.

\begin{figure}[h]
\includegraphics[trim = 20 0 30 20, clip, width = 1\linewidth]{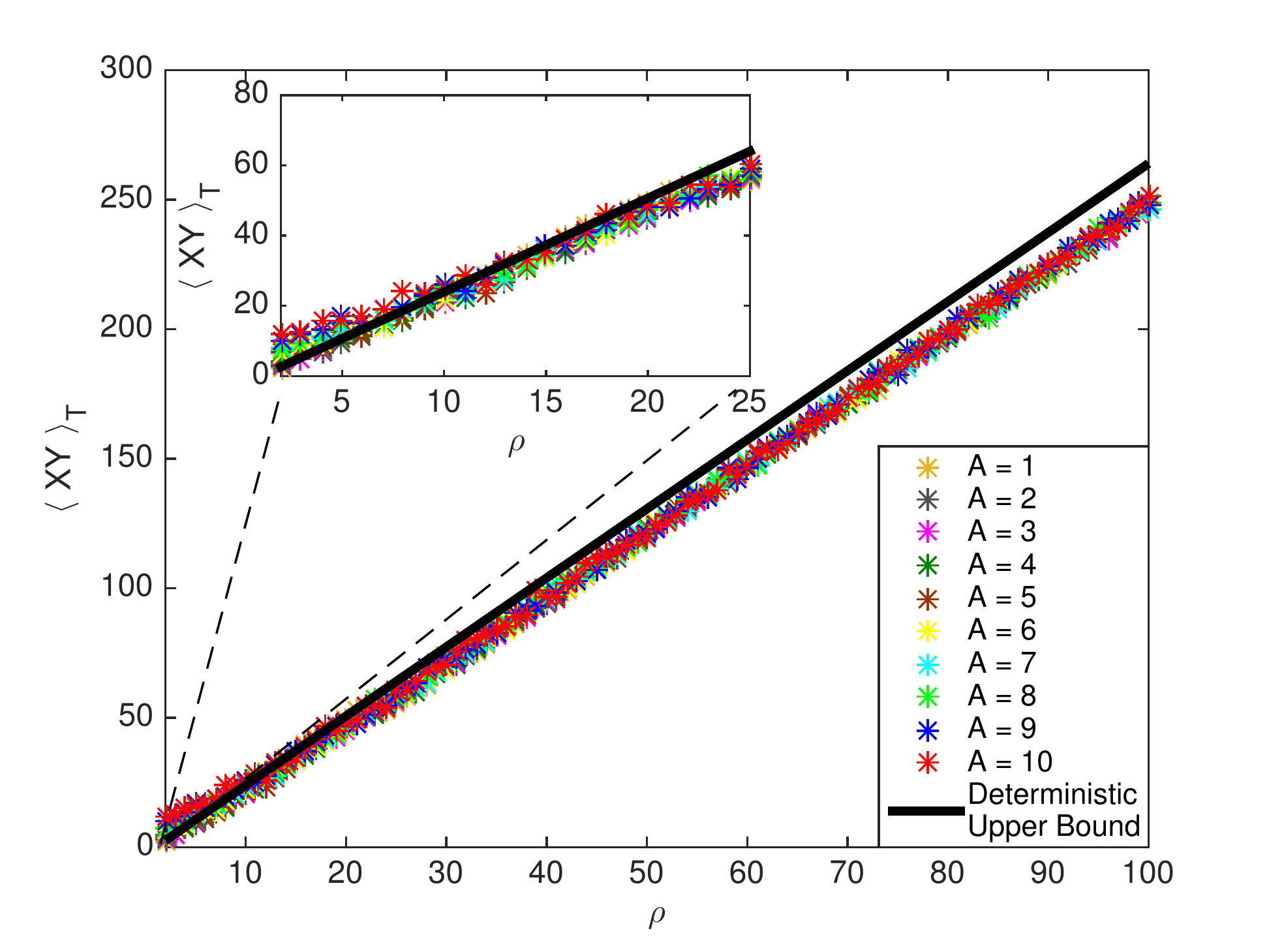}
\vspace{-0.3in}
\caption{$\overline{\lim_{T \rightarrow \infty}} \left \langle XY \right \rangle _{T}$, the transport from a single realization of the stochastic Lorenz attractor (equation  \ref{eq:StoLorenz}),  as a function of $\rho$ and noise amplitude $A$, with the solid black line showing the deterministic upper bound \cite{Doering15}. The inset shows the increased transport for $\rho$ near the transition to chaos; $\rho_c = 24.74$, beyond which the solutions cross below the deterministic upper bound.} 
\label{fig:XY}
\vspace{-0.2in}
\end{figure}

\section{Results \& Interpretation \label{sec:Results}}

It is often the case that for the nonlinear dynamical systems found in nature, we only have a single time series.  Thus, it is a natural question to ask about the properties of the stochastic upper bound both for the ensemble average {\em and} for a small number of realizations.  Whereas deterministic chaos acts to decrease the transport in the system \cite{Doering15}, here we find that it can also be indistinguishable from noise.   In Fig. \ref{fig:XY} we show individual realizations (one for each of 10 amplitudes $A$) of the transport as a function of $\rho$.  These exhibit two important features.  (1) For $\rho$ below the deterministic transition to chaos ($\rho_c = 24.74$), the solutions transcend the deterministic upper bound (DUB), whereas for $\rho$ above $\rho_c$ they cross below it.  (2) Independent of $\rho$, there is no systematic dependence of the solutions on the noise amplitude.  Taken together these features show that the impact of noise differs substantially depending on whether the deterministic dynamics is chaotic or non-chaotic.  Clearly, the role of noise is indistinguishable from the role of chaotic dynamics and in individual realizations a given noise amplitude couples with various unstable periodic orbits, which we discuss in more detail below.  

\begin{figure}[h]
\includegraphics[trim = 30 0 30 20, clip, width = 1\linewidth]{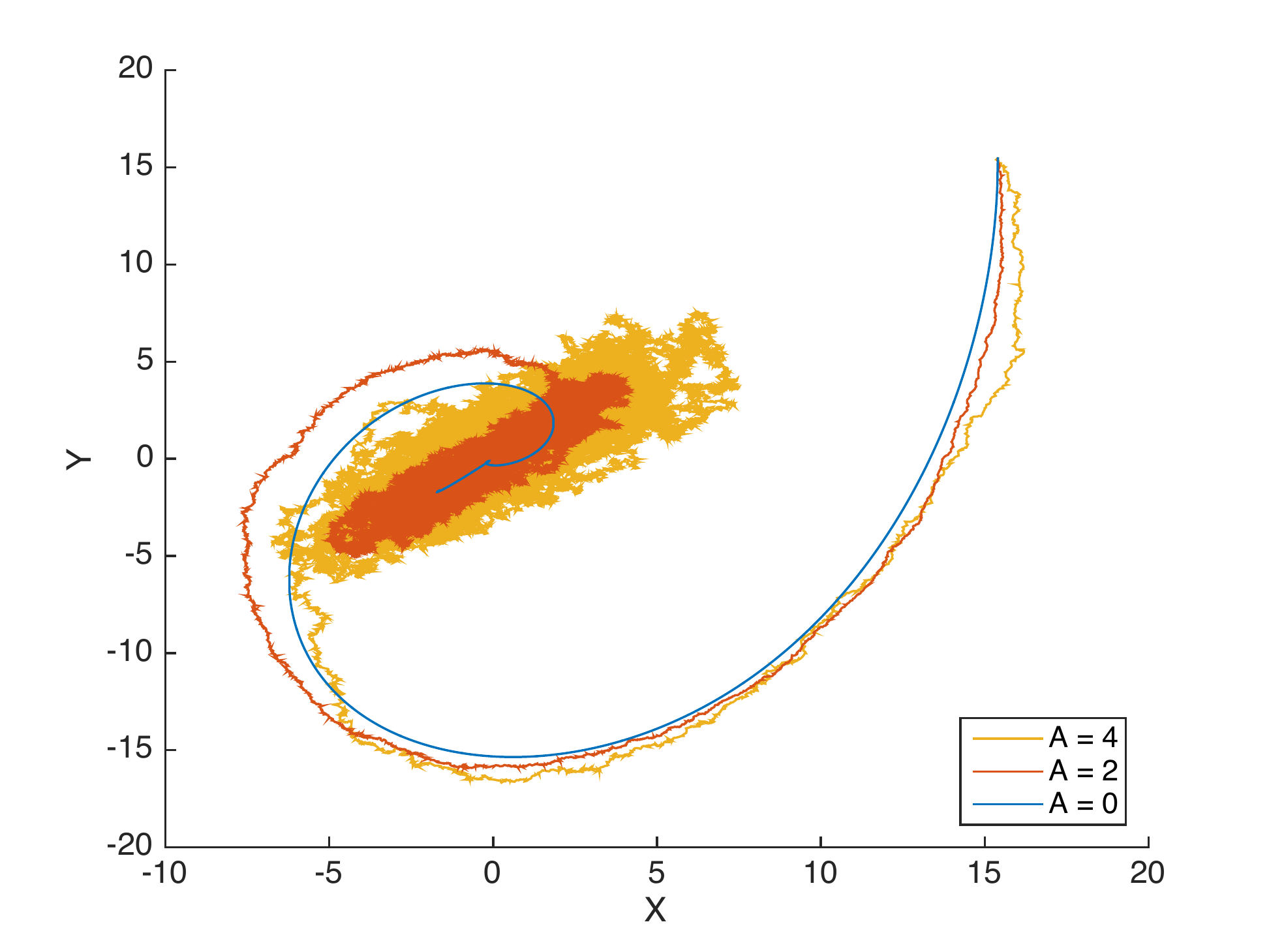}
\vspace{-0.3in}
\caption{XY-space of the stochastic Lorenz attractor (equation \ref{eq:StoLorenz}), for three different noise amplitudes. The diameter of the attractor increases with noise amplitude.} 
\label{fig:rhoXY}
\vspace{-0.2in}
\end{figure}

The analytical solution from equation \ref{eq:XYZleq} and the numerical solution (taking the ensemble average in equation \ref{eq:SXYZlimit}) of the stochastic upper bound (SUB) are shown in Fig. \ref{fig:SUBXY}. 
Firstly, we see the increase in the SUB as the noise amplitude $A$ increases. Secondly, for fixed amplitude and values of $\rho < \rho_c$, the origin of the increase in the SUB are the two terms proportional to $1/\rho$. As $\rho$ increases and $A$ decreases the SUB converges to the DUB from above. The increase with $A$ at low $\rho$ is a reflection of the dependence of the ``diameter'' of the system in the $X-Y$ plane, other parameters being held constant. Because the diameter decreases as $\rho$ decreases, the relative influence of $A$ on the SUB is larger.  We show this for $\rho = 2$ in Fig. \ref{fig:rhoXY}.  However, we note that, using a different method Fantuzzi and Goluskin (pers. comm.) find a SUB that does not exhibit the low $\rho$ divergence, asymptotes to our bound for $\rho$ in the region of typical interest, and also recovers the DUB in the appropriate limit.  

\begin{figure}[h]
\includegraphics[trim = 20 0 30 20, clip, width = 1\linewidth]{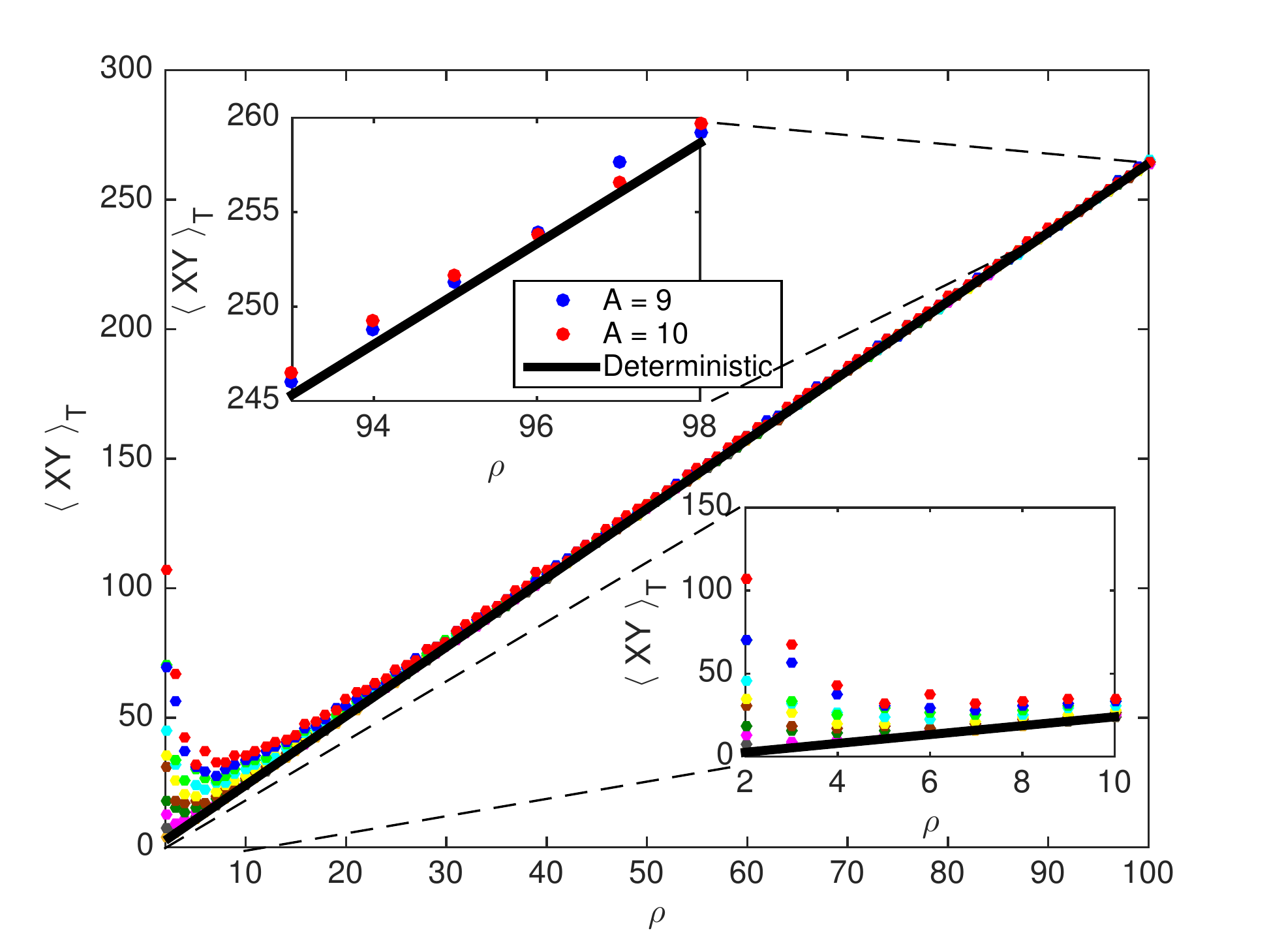}
\vspace{-0.3in}
\caption{The stochastic upper-bound (circles) from Eq. \ref{eq:SXYZlimit} as a function of $\rho$ and noise amplitude $A$.  The solid black line is the DUB \cite{Doering15}. The bottom inset shows that the SUB is a monontoic function of $A$ in the non-chaotic regime ($\rho < \rho_c$). The top inset shows the oscillations of the SUB between noise amplitudes $A = 9,10$ in the chaotic regime ($\rho < \rho_c$).} 
\label{fig:stoUBinset}
\vspace{-0.2in}
\end{figure}

A more detailed view of the SUB from Eq. \ref{eq:SXYZlimit} is shown in Fig \ref{fig:stoUBinset}. The lower right inset shows that for $\rho < \rho_c$ the SUB
is a monotonic function of the noise amplitude, as one would intuitively expect from Fig. \ref{fig:rhoXY}. However, as the system enters the chaotic regime, this monotonicity is lost to reveal an oscillation with amplitude, as shown in the upper inset of Fig. \ref{fig:stoUBinset} for noise amplitudes $A = 9$ and $A = 10$. This oscillatory behavior is due to the coupling of noise with chaotic orbits that, depending on the amplitude, can result in different residence times of a trajectory in different orbits.   Indeed, although for $\rho < \rho_c$, the realization to realization stochastic upper bounds are consistent, this is not the case in the chaotic regime due to the coupling between the noise and the chaotic orbits.  In consequence, each realization results in a slightly different SUB and hence the bound is not strict; it has a diffuseness that depends on the noise amplitude.  This combined effect of noise and the exponential divergence property of chaos allows the noise to perturb the stochastic system into a different orbit in each realization.


\begin{figure}[h]
\includegraphics[trim = 30 180 30 180, clip, width = 1.1\linewidth]{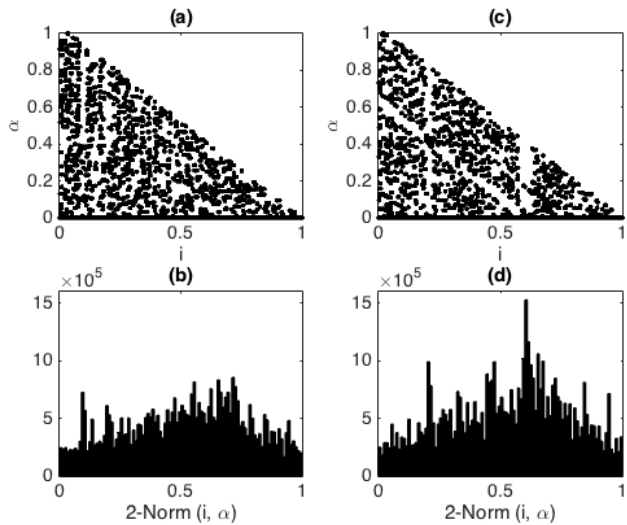}
\vspace{-0.3in}
\caption{Scaled close returns plots (a,c) and the corresponding histograms of the $2-$norm of these (b,d) for the stochastic Lorenz attractor (equation  \ref{eq:StoLorenz}) with $\rho = 97$, where the horizontal segments represent the UPOs in the system when they close up when embedded in the phase space of the attractor. The scaled index in the time series is $i$, and $\alpha$ is the scaled period of the UPO. $(a,b)~A = 9$. $(c,d)~A = 10$. } 
\label{fig:CR_Histr97}
\vspace{-0.2in}
\end{figure}

The {\em close returns plot} of \citet{Mindlin92} can be used to extract unstable periodic orbits (UPO) from a chaotic time series.
Thus, to demonstrate the coupling between noise and chaos discussed above in a different manner, we show the close returns plot for the stochastic Lorenz attractor (equation  \ref{eq:StoLorenz}) for $A = 9, 10$ and $\rho = 97$ in Fig. \ref{fig:CR_Histr97} $(a,c)$.  These plots help us distinguish between noise and chaos. Whereas in a noisy system the points are more diffuse, in a chaotic system they are more structured, with continuous straight lines defining the UPOs. 


To further quantify this structure, in Fig. \ref{fig:CR_Histr97} $(b,d)$ we plot the histogram of the $2-$norm of the points from the close returns plot. The peaks in the histogram show the UPOs in the system, corresponding to the lines in the close returns plot. Whereas the histogram in Fig. \ref{fig:CR_Histr97}$(b)$ has a broad-band structure, that in Fig. \ref{fig:CR_Histr97}$(d)$ reveals prominent peaks at different norms.

\section{Summary} 

We calculated the stochastic upper bounds of the heat for the Lorenz equations using an extension of the background method of \citet{Doering15} used in the deterministic system.  Whilst one might have expected that the stochastic upper bounds transcend their deterministic counterpart of \cite{Doering15}, their variation with noise amplitude exhibits rich behavior.  In the non-chaotic regime the upper bounds increase monotonically with noise amplitude.  However, in the chaotic regime this 
monotonicity depends on the number of realizations in the ensemble; at a particular Rayleigh number the bound may increase or decrease with noise amplitude.  
The origin of this behavior is the coupling between the noise and unstable periodic orbits, the degree of which depends on the degree to which the ensemble represents the ergodic set.  This is confirmed by examining the close returns plots of the full solutions to the stochastic equations.  These solutions also demonstrate that the effect of noise is equivalent to the effect of chaos for a wide range of noise amplitude.  Finally, we note that although in \emph{It\^{o}}-calculus the analytic bound (equation \ref{eq:XYZleq}) relies on vanishing noise correlations ($\langle X\xi_1 \rangle = \langle Y\xi_2 \rangle = \langle \lambda \xi_3 \rangle = 0$), numerically such correlations never completely vanish \cite[e.g.,][]{Ito:1984}\footnote{To test time convergence in the It\^{o} Case  we take the time average of these noise correlations  for turnover times of 50 and 1550, and find that the magnitude of the correlations are of the same order; the correlations change from $\sim$ 0.05 (50 turnover times) to $\sim$ 0.015 (1550 turnover times). 
For the ensemble average convergence, with a single member the noise correlations are $\sim$ 0.05 (with 50 turnover times), and for 1000 members (with 50 turnover times) these are $\sim$ 0.001.  Thus, both time and ensemble averages converge slowly with the latter being slightly superior.}, for as the size and diffuseness of stochastic attractor continually increases, the extent to which ensemble average reaches the ergodic set remains a concept rather than a practical reality.  

%
%

\acknowledgments

SA and JSW acknowledge NASA Grant NNH13ZDA001N-CRYO for support.  JSW acknowledges the Swedish Research Council and a Royal Society Wolfson Research Merit Award for support.  This work was completed whilst the authors were at the 2015 Geophysical Fluid Dynamics Summer Study Program at the Woods Hole Oceanographic Institution, which is supported by the National Science Foundation and the Office of Naval Research.  We thank C.R. Doering, G. Fantuzzi, D. Goluskin and A. Souza for feedback.


%

\end{document}